\documentclass[aps,prl,preprint,tightenlines,superscriptaddress,showpacs,byrevtex,preprintnumbers,amsmath,amssymb]{revtex4}

\usepackage{graphicx}
\usepackage{dcolumn}
\usepackage{bm}

\newcommand{\GeV}{\ensuremath{\text{GeV}}}
\newcommand{\MeV}{\ensuremath{\text{MeV}}}
\newcommand{\bgs}{\ensuremath{B\to X_s\gamma}}
\newcommand{\ifb}{\ensuremath{\,\rm fb^{-1}}}

\begin{document}

\begin{tabular}{l}
\includegraphics[scale=0.20]{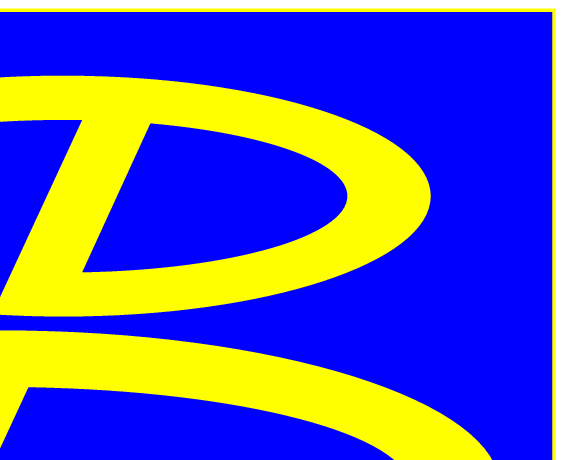} \\
\end{tabular}

\preprint{\vbox{ \hbox{   }
                        \hbox{Belle Preprint 2009-16}
                        \hbox{KEK Preprint   2009-14}
}}

\title{
  Measurement of Inclusive Radiative B-meson Decays with a Photon
  Energy Threshold of 1.7 GeV
}

\begin{abstract}
Using $605\ifb$ of data collected at the $\Upsilon(4S)$ resonance we present 
a measurement of the inclusive radiative $B$-meson decay channel,  
$B\to X_s \gamma$. 
For the lower photon energy thresholds of $1.7, 1.8, 1.9$ and $2.0\,\GeV$, 
as defined 
in the rest frame of the $B$-meson, we measure the partial branching fraction 
and the mean and variance of the photon energy spectrum. 
At the $1.7\,\GeV$ threshold we obtain the partial branching fraction 
$\mathrm{BF}\left( B\to X_s \gamma \right) = \left( 3.45 \pm 0.15
\pm 0.40 \right)\times 10^{-4}$, 
where the errors are statistical and systematic.

\end{abstract}


\affiliation{Budker Institute of Nuclear Physics, Novosibirsk}
\affiliation{University of Cincinnati, Cincinnati, Ohio 45221}
\affiliation{T. Ko\'{s}ciuszko Cracow University of Technology, Krakow}
\affiliation{The Graduate University for Advanced Studies, Hayama}
\affiliation{Hanyang University, Seoul}
\affiliation{University of Hawaii, Honolulu, Hawaii 96822}
\affiliation{High Energy Accelerator Research Organization (KEK), Tsukuba}
\affiliation{Institute of High Energy Physics, Chinese Academy of Sciences, Beijing}
\affiliation{Institute of High Energy Physics, Vienna}
\affiliation{Institute of High Energy Physics, Protvino}
\affiliation{Institute for Theoretical and Experimental Physics, Moscow}
\affiliation{J. Stefan Institute, Ljubljana}
\affiliation{Kanagawa University, Yokohama}
\affiliation{Institut f\"ur Experimentelle Kernphysik, Universit\"at Karlsruhe, Karlsruhe}
\affiliation{Korea University, Seoul}
\affiliation{Kyungpook National University, Taegu}
\affiliation{\'Ecole Polytechnique F\'ed\'erale de Lausanne (EPFL), Lausanne}
\affiliation{Faculty of Mathematics and Physics, University of Ljubljana, Ljubljana}
\affiliation{University of Maribor, Maribor}
\affiliation{University of Melbourne, School of Physics, Victoria 3010}
\affiliation{Nagoya University, Nagoya}
\affiliation{Nara Women's University, Nara}
\affiliation{National Central University, Chung-li}
\affiliation{National United University, Miao Li}
\affiliation{Department of Physics, National Taiwan University, Taipei}
\affiliation{H. Niewodniczanski Institute of Nuclear Physics, Krakow}
\affiliation{Nippon Dental University, Niigata}
\affiliation{Niigata University, Niigata}
\affiliation{Novosibirsk State University, Novosibirsk}
\affiliation{Osaka City University, Osaka}
\affiliation{Panjab University, Chandigarh}
\affiliation{Saga University, Saga}
\affiliation{University of Science and Technology of China, Hefei}
\affiliation{Seoul National University, Seoul}
\affiliation{Sungkyunkwan University, Suwon}
\affiliation{University of Sydney, Sydney, New South Wales}
\affiliation{Toho University, Funabashi}
\affiliation{Tohoku Gakuin University, Tagajo}
\affiliation{Tohoku University, Sendai}
\affiliation{Department of Physics, University of Tokyo, Tokyo}
\affiliation{Tokyo Metropolitan University, Tokyo}
\affiliation{Tokyo University of Agriculture and Technology, Tokyo}
\affiliation{IPNAS, Virginia Polytechnic Institute and State University, Blacksburg, Virginia 24061}
\affiliation{Yonsei University, Seoul}
  \author{A.~Limosani}\affiliation{University of Melbourne, School of Physics, Victoria 3010} 
  \author{H.~Aihara}\affiliation{Department of Physics, University of Tokyo, Tokyo} 
\author{K.~Arinstein}\affiliation{Budker Institute of Nuclear Physics, Novosibirsk}\affiliation{Novosibirsk State University, Novosibirsk} 
  \author{T.~Aushev}\affiliation{\'Ecole Polytechnique F\'ed\'erale de Lausanne (EPFL), Lausanne}\affiliation{Institute for Theoretical and Experimental Physics, Moscow} 
  \author{A.~M.~Bakich}\affiliation{University of Sydney, Sydney, New South Wales} 
  \author{V.~Balagura}\affiliation{Institute for Theoretical and Experimental Physics, Moscow} 
  \author{E.~Barberio}\affiliation{University of Melbourne, School of Physics, Victoria 3010} 
  \author{A.~Bay}\affiliation{\'Ecole Polytechnique F\'ed\'erale de Lausanne (EPFL), Lausanne} 
  \author{K.~Belous}\affiliation{Institute of High Energy Physics, Protvino} 
  \author{M.~Bischofberger}\affiliation{Nara Women's University, Nara} 
% ...
\collaboration{The Belle Collaboration}

\maketitle


Radiative $B$-meson decays $B\to X_s\gamma$ may offer a view of phenomena beyond the
Standard Model of Particle Physics (SM).
These decays proceed via a flavor changing neutral current
process; yet to be discovered hypothetical particles, e.g. in the
Minimal Supersymmetric Standard Model~\cite{Bertolini:1990if} or left-right symmetric
model~\cite{Cho:1993zb}, may contribute and cause a sizeable deviation from the
branching fraction (BF) predicted by the SM.  The BF prediction,
$\left(3.15 \pm 0.23\right)\times10^{-4}$~\cite{Misiak:2006zs}, and the average
experimental value, $\left(3.56\pm0.26\right)\times10^{-4}$~\cite{Amsler:2008zzb}, are in marginal
agreement.  A measurement with improved precision provides a more
stringent test and gives stronger constraints on models beyond the SM.

The photon energy spectrum is a direct probe of the $b$-quark's mass
and Fermi motion and therefore provides information needed to extrapolate the
photon spectrum below the lower photon energy threshold~\cite{Buchmuller:2005zv}, as well as that for the
extraction of the SM parameters  $|V_{cb}|$ and $|V_{ub}|$ from inclusive semileptonic
$B$ decays~\cite{Barberio:2007cr}. The lower the threshold
the smaller are their model uncertainties~\cite{Bigi:2002qq}.

Belle has previously measured \bgs\ with $5.8\ifb$ and $140\ifb$ of data using
semi-inclusive~\cite{Abe:2001hk} and fully
inclusive approaches~\cite{Koppenburg:2004fz}, respectively. Here we
present a new measurement,
based on a larger data set and with
significant improvements. We cover more of the spectrum by extending the photon energy range from
$1.8\,\GeV$ down to $1.7\,\GeV$, as measured in the $B$-meson rest frame.
CLEO~\cite{Chen:2001fja} and BaBar~\cite{Aubert:2005cua} reported measurements at
lower thresholds of $2.0\,\GeV$ and $1.9\,\GeV$, respectively.

We use data collected by the Belle detector at the KEKB
asymmetric-energy $e^+e^-$ storage ring~\cite{Kurokawa:2001nw}. The data consists of
a sample of $605\ifb$ taken on the $\Upsilon(4S)$ resonance
corresponding to $657 \times10^6$ $B\bar{B}$ pairs. Another
$68\ifb$ sample has been taken at an energy $60\,\MeV$ below the resonance (off-resonance).
The Belle detector is a large-solid-angle magnetic spectrometer
described in detail elsewhere~\cite{Abashian:2000cg}.
The main component relevant for this analysis is the 
electromagnetic calorimeter (ECL) made of 
$16.2$ radiation lengths long CsI(Tl) crystals. The photon energy resolution
is about 2\% for the energy range explored in this analysis.

We extract the signal \bgs\ spectrum by collecting all 
high-energy photons, vetoing those originating from $\pi^0$ and $\eta$ 
decays to two photons. The contribution from non-$B\bar{B}$ processes, referred to as continuum background, 
mainly $e^+e^-\to q\bar{q}$ ($q=u,d,s,c$) events, is subtracted using
the off-resonance sample. The remaining 
backgrounds from $B\bar{B}$ events are subtracted using 
Monte-Carlo (MC) simulated distributions normalized using data control samples.

Photon candidates are selected from ECL clusters of $5\times5$ crystals
in the barrel region, $\cos\theta_\gamma \in [-0.35,0.70]$, where $\theta_\gamma$ 
is the polar angle with respect to the beam axis, subtended from the
direction opposite the positron beam. 
They are required to have energies 
$E^\mathrm{c.m.s.}_\gamma$ larger than $1.4\,\GeV$, as measured in the
center-of-mass system of the $\Upsilon(4S)$(c.m.s.). Further selection
criteria, the same as those applied in Ref.~\cite{Koppenburg:2004fz},
are applied to ensure that clusters are isolated in the ECL and
cannot be matched to tracks reconstructed in the drift chamber.

Owing to increased instantaneous luminosity delivered by KEKB, there
is a non-negligible background due to the overlap of hadronic events
with energy deposits left in the calorimeter by earlier 
QED interactions (mainly Bhabha scattering). Such composite events are completely separated from the signal 
using timing information for calorimeter clusters 
associated with the candidate photons.
The cluster timing information is stored in the raw data,
and is available in the reduced format used for
analysis of data processed after the summer of 2004.
Our data set is divided into  $254\ifb$ and $351\ifb$  samples
that correspond to sub-samples without and with cluster timing information, respectively.
In the second data set photons that are off-time are rejected with a signal inefficiency of $0.2\%$.  In
addition, for both sets, we veto any event that contains an ECL cluster with energy
exceeding $1$ GeV within a cone of 0.2 radians in the
direction opposite our photon candidate in the c.m.s. We employ the
same criteria as those used in Ref.~\cite{Koppenburg:2004fz} to
veto candidate photons from $\pi^0$ and $\eta$ decays to two photons.

The analysis proceeds in two different streams, with a lepton tag (LT) and without
(MAIN), resulting in final samples that give similar
sensitivity to the signal while being largely statistically
independent. The lepton tag is employed to suppress continuum background, since the
presence of a high momentum lepton is more likely to originate from a
primary semileptonic decay of the second $B$-meson in signal events.
For an event to be accepted in the LT stream, it must
contain a well identified electron or muon, with momentum between $1.26$
GeV/$c$ and $2.20$ GeV/$c$, at an angle $\theta_\ell$ to the candidate
photon such that $\cos\theta_\ell \in [-0.67,+0.87]$, where all
measurements are in the c.m.s.
The event selection criteria employed to
reduce the contribution from continuum events in the MAIN stream are
the same as applied in our previous
analysis, namely two Fisher discriminants, relying on energy flow and
event-shape variables. Only the second of these discriminants is used
in the LT stream with the corresponding coefficients calculated
for the lepton tag.

To optimize these selection criteria, we use MC
simulated~\cite{Brun:1984} large samples of $B\bar{B}$,
$q\bar{q}$ and signal. The signal MC events are generated as a
mixture of exclusive $B\to K^*\gamma$ and inclusive $B\to X_s\gamma$
components using EvtGen~\cite{Lange:2001uf}. The inclusive component
$X_s$ is first generated as a $s\bar{u}$ or
$s\bar{d}$ state with spin-1, and then hadronized by JETSET~\cite{Sjostrand:2000wi}.
The relative weights of the two components are chosen to match the
world average branching fractions for $B\to K^*\gamma$ and $B\to X_s\gamma$~\cite{Amsler:2008zzb}.
To improve the understanding of the photon energy spectrum 
at low energies, the selection criteria are
optimized in
the energy bin $1.8\,\GeV<E^\mathrm{c.m.s.}_\gamma<1.9\,\GeV$.

After these selection criteria we observe $41.1\times 10^5$ ($24.6\times
10^4$) and $3.5\times 10^5$ ($0.9\times 10^4$) 
photon candidates in the MAIN (LT) stream of the on- and
off-resonance data samples, respectively.
The spectrum measured in off-resonance data is scaled by the ratio of
the on- to off-resonance luminosity and subtracted.
We apply corrections to the yield and energy of candidates derived from
the off-resonance sample to account for the effects of the
$60\,\MeV$ ($0.5$\%) energy difference: a lower cross-section
and, on average, lower multiplicity and energy of photon candidates.

Beam background is estimated using a sample of randomly triggered 
events that is added to the $B\bar{B}$ MC. The remnant beam background
left after subtraction of non-$B\bar{B}$ background
is reduced to a negligible level after the application of the off-time veto.
In the sample of data where the veto is unavailable, we scale the
background according to a comparison of yields between MC and data for
high energy ($E^\mathrm{c.m.s.}_\gamma>2.8$ GeV) photon candidates found in the
endcaps of the ECL. This sample after off-resonance subtraction is a clean
sample of ECL clusters from beam backgrounds. The ratio of beam
background data to MC is found to be $1.16\pm 0.04$ in this sample. 

From the on-resonance spectrum after continuum background subtraction we subtract
backgrounds from $B$ decays.
We divide the background into six categories, with relative
contributions after selection in the $1.7 \mathrm{\,GeV}
<E^\mathrm{c.m.s.}_\gamma <2.8 \mathrm{\,GeV}$
range as follows for the two streams (MAIN/LT):
(i) photons from $\pi^0\to\gamma\gamma$ (47.4\%/48.0\%);
(ii) photons from $\eta\to\gamma\gamma$ (16.3\%/16.0\%);
(iii) other real photons, mainly from
decays of $\omega$, $\eta'$, and $J/\psi$ mesons, 
and bremsstrahlung,
including the short distance radiative
correction~\cite{Barberio:1993qi}
(8.1\%/8.9\%);
(iv) ECL clusters not due to single photons, mainly from $K^0_L$'s and
  $\bar{n}$'s (1.7\%/1.6\%);
(v) electrons 
  misidentified as photons (6.1\%/3.3\%) and;
  (vi) beam background (1.3\%/2.6\%). 
  The signal fractions are 19.1\% and
19.6\%, respectively. 

For all of the selection criteria and for each background
category we determine the $E^\mathrm{c.m.s.}_\gamma$-dependent
selection efficiency in off-resonance subtracted data and MC using
appropriate control samples as described in
Ref.~\cite{Koppenburg:2004fz}.
The ratios of data and MC efficiencies versus
$E^\mathrm{c.m.s.}_\gamma$ are fitted using low-order polynomials,
which are used to scale the background MC. 
Most are found to be statistically compatible with unity.
An exception is the effect of the selection requirements on ECL
clusters produced by hadrons: specifically, the requirement that $95$\% of the energy
be deposited in the central nine cells of the $5\times5$
cluster, which is poorly modelled in the MC simulation. The correction
doubles the yield of the hadron background~\cite{Koppenburg:2004fz}.

Each background yield, after having been 
properly scaled by the described procedures, is subtracted from 
the data spectrum. The spectra for the MAIN and LT streams
are shown in Fig.~\ref{fig:signalave}(a) and
Fig.~\ref{fig:signalave}(b), respectively.
The photon energy ranges $1.4\mbox{--}1.7$ GeV and
$2.8\mbox{--}4.0$ GeV were chosen a priori as control regions to test the integrity of the background
subtraction since in the low energy region the little signal expected is negligible with
respect to the uncertainty on the background, and
no signal is possible in the high energy region above the kinematic limit.
The yields in the high energy region
are $1245 \pm 4349$ and $292 \pm 410$ candidates
in the MAIN and LT streams, respectively, while corresponding yields
in the low energy region are $-1629 \pm 3071$ and $-745 \pm 623$, respectively.

To obtain the true spectrum we correct the
raw spectrum in the energy range $1.4\mbox{--}2.8$ GeV using
a three-step procedure: (i) divide by the probability of a photon
  candidate satisfying selection criteria given a
  cluster has been selected in the ECL;
(ii) perform an unfolding procedure based on the Singular Value
Decomposition (SVD) algorithm~\cite{Hocker:1995kb}, which maps
the spectrum from measured energy to true
energy thereby undoing the distortion caused by the ECL; and
(iii) divide by the probability that a photon originating at the
interaction point is detected in the ECL.
The average efficiency of step (i) over the entire range is
15\% (2.5\%) in the MAIN (LT) stream. The average efficiency of
step (iii) is about 80\%. Step (ii) eliminates the need to perform
corrections for the effect of the ECL resolution on the
moments, as was done in Ref.~\cite{Schwanda:2008kw},
and thereby significantly
reduces the large uncertainty due to model dependence.
The unfolding matrix, derived from the
signal MC sample, is calibrated to data using the results of a study
of a clean photon sample from radiative $\mu$-pair events. 

A weighted average taking into account the correlation of the MAIN and
LT stream spectra is performed after step (i). At this stage the averaging
procedure is substantially simplified
since there is no statistical correlation between yields in different
energy bins. As an example of the cross correlation between the MAIN and
LT streams, in the energy bin $2.00-2.05$ GeV, there are 116517 (9834)
and 6769 (246) photon candidates in the
on-(off-)resonance sample, in the MAIN and LT streams, respectively,
of which, 3815 (72) are common to both streams. We find the
covariance between the MAIN and LT signal yields is dominated by the
overlap of candidates from the off-resonance sample,
which is small relative to the individual variances of the MAIN and LT
signal yields. This results in a statistical error on the average just
above that which is obtained if no statistical correlation is
assumed. The spectrum derived from the average of MAIN and LT stream
spectra before unfolding is shown in
Fig~\ref{fig:signalave}(c). 
\begin{figure*}
  \begin{tabular}{ccc}
    \hspace{-5mm} \includegraphics[scale=0.25]{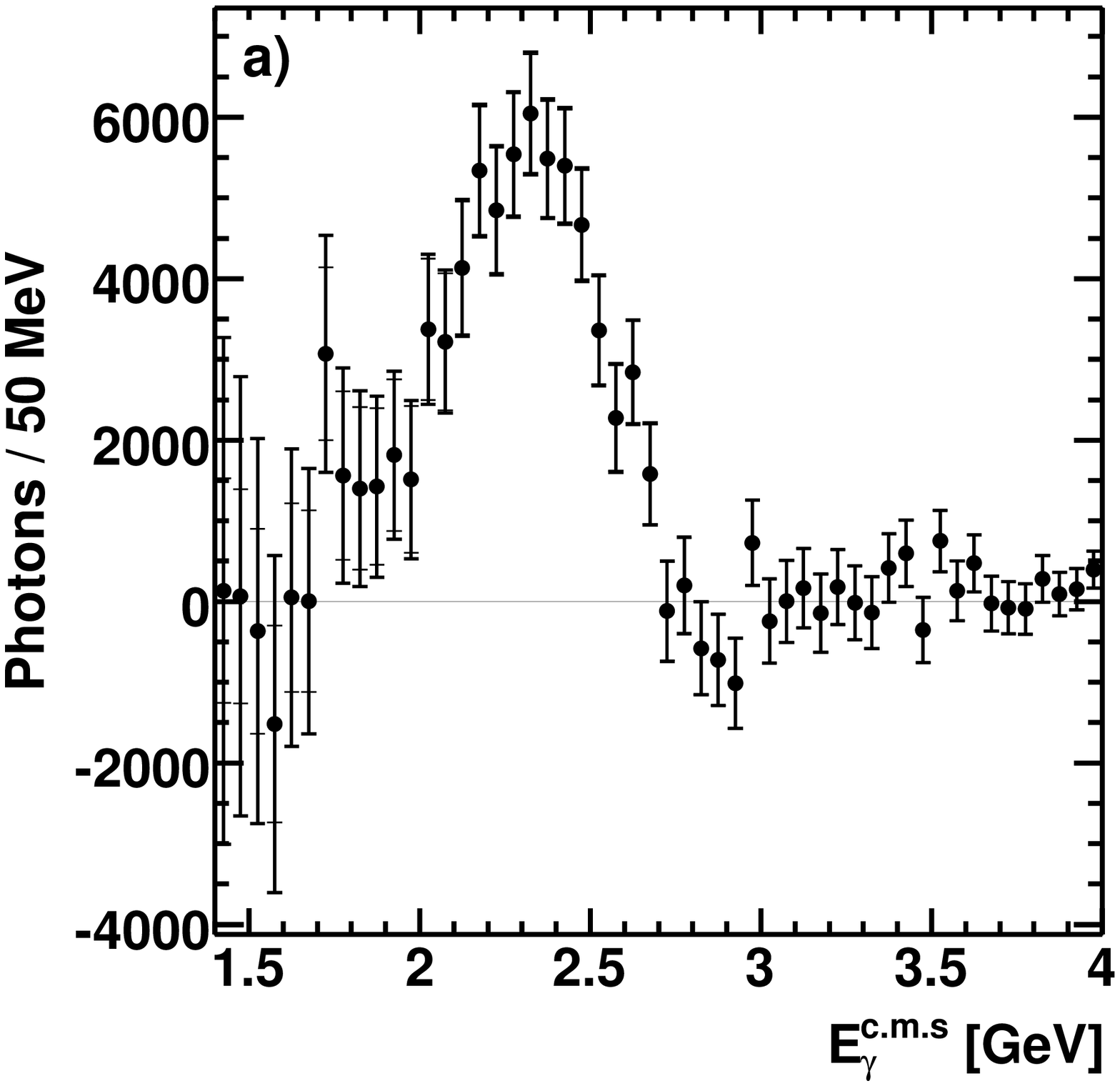} &
    \includegraphics[scale=0.25]{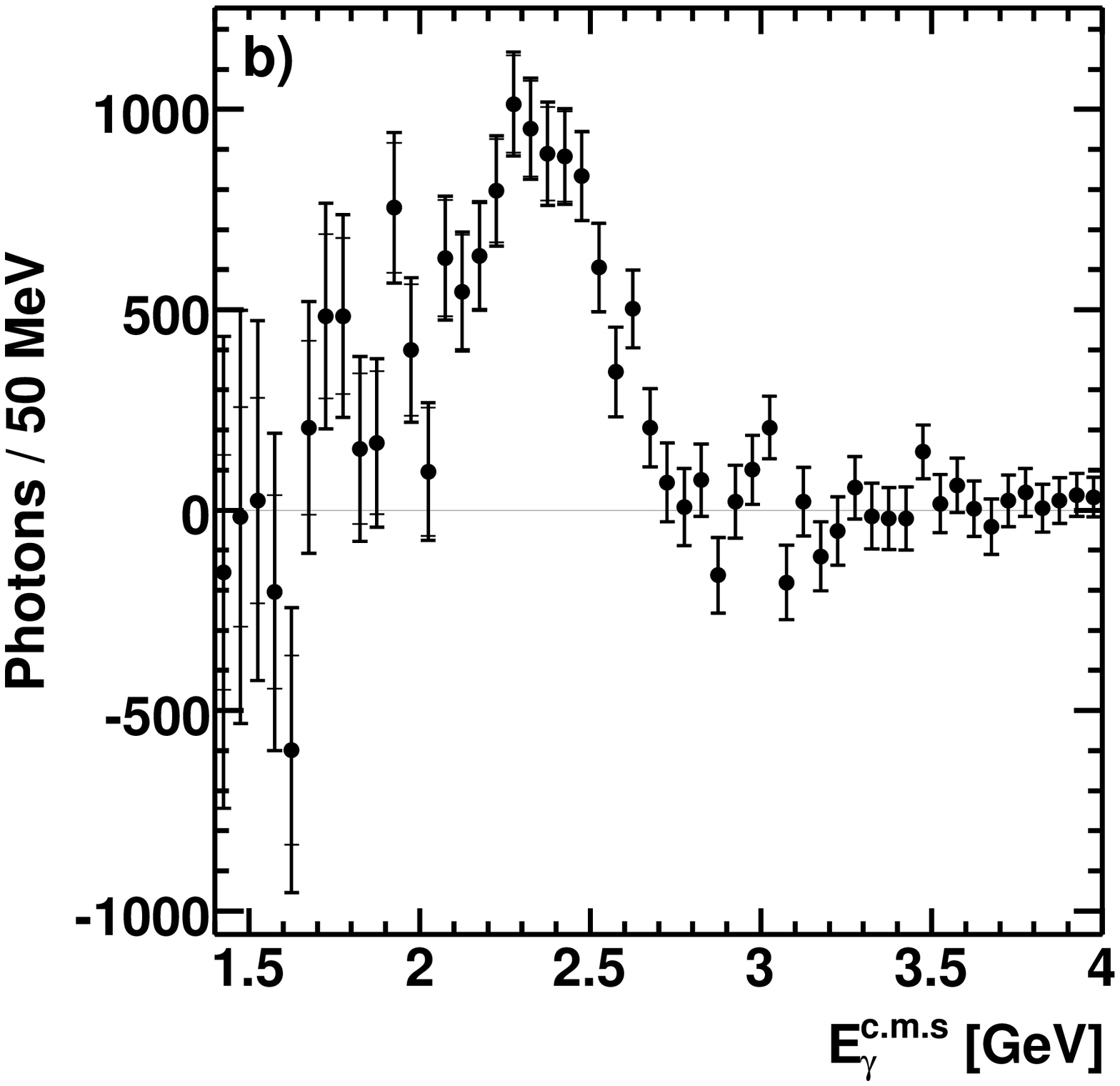} &
    \includegraphics[scale=0.25]{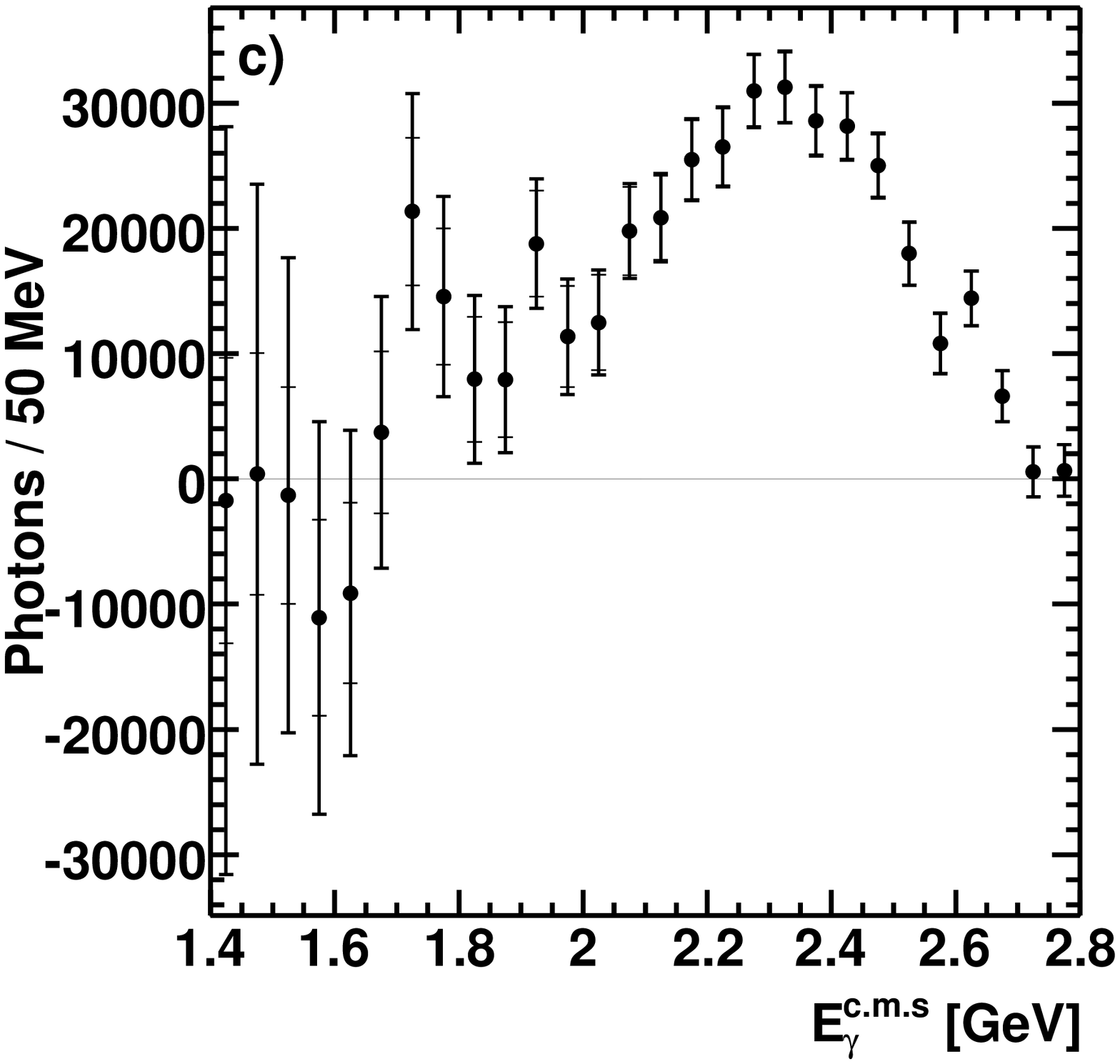} \\
  \end{tabular}
\caption{\label{fig:signalave}
     The extracted photon energy spectrum of $B\to X_{s,d}\gamma$ in
    the (a) MAIN and (b) LT stream before any correction for
    signal acceptance is applied; and c) displays their average after
    correction by the selection efficiency. The two error bars for each point show the statistical
and the total error. The total error is a sum in quadrature of 
the statistical and systematic errors, where the latter are correlated
between bins. The LT and MAIN streams refer to the set of selection
     criteria that do and do not include the lepton tag criterion, respectively.
}
\end{figure*}

Our analysis procedure does not distinguish
between $B\to X_s \gamma$ and $B \to X_d \gamma$.  We subtract the
contribution of the latter from all partial branching fraction measurements by
assuming the ratio of the branching fractions to be $R_{d/s} =
(4.5\pm0.3)\%$~\cite{Hurth:2003dk,Charles:2004jd}, and thereby assume the shape of the
corresponding photon energy spectra to be equivalent. Employing
other models for the $B \to X_d \gamma$ photon energy
spectrum has a negligible impact on the measured branching fractions and
moments of $B \to X_s \gamma$.

To derive the measurements in the rest frame of the $B$-meson we
calculate boost corrections using a MC simulation. The
corrections are calculated from differences between the spectra in the
$B$-meson and c.m.s. frames. The simulation takes into account the
energy of the $B$-meson and its angular distribution in the c.m.s.

Systematic uncertainties are calculated from a number of sources, as
given by the numbered list in Table~\ref{Tab:AVERAGEsummary}.
We vary the number of $B\bar{B}$ pairs, the on-resonance to off-resonance ratio of integrated
luminosities and the correction factors applied to the off-resonance photon
candidates and assign the observed variation
as the systematic error associated with continuum subtraction (1).
The parameters of the correction functions applied to the MC to
calibrate for the effect of selection criteria (2) and those applied to the
the $\pi^0$ and
$\eta$ yields (3)
are varied taking into account their correlations. 
As we do not measure the yields of photons from sources other than 
$\pi^0$'s and $\eta$'s in $B\bar{B}$ events, we independently vary the 
expected yields of these additional sources by $\pm 20$\% (4).
We vary the corrections applied to beam background data according to their uncertainties (5).
For the uncertainties related to the unfolding
procedure, we vary the value of the regularization parameter of the SVD algorithm (6).
We compare the results from five signal
models~\cite{Kagan:1998ym}
with corresponding model parameters derived
from fits to the signal spectrum derived from the MAIN
stream shown in Fig~\ref{fig:signalave}(a). We assign the maximum
deviation from the Kagan-Neubert model as the
uncertainty (7). The errors associated to the measurement of the 
photon energy resolution and photon detection efficiency in radiative $\mu$-pair events are varied (8,9), 
where the former has an uncertainty of 1\%. 
To account for the higher multiplicity hadronic
environment of $B\bar{B}$ decays and secondary effects in the estimation of
photons from $B\bar{B}$, we assign twice ($\pm 2\sigma$) the variation as
the associated systematic error (9). We find the signal yield as derived after acceptance
correction is susceptible to the statistical fluctuations evident in the lower
energy region of the photon energy spectrum measured in the LT stream, which propagate from the 
off-resonance sample. 

In the photon energy range from $1.7\,\GeV$ to $2.8\,\GeV$, as measured in
the $B$-meson rest frame,  
we obtain the partial branching fraction: $
  \mathrm{BF}\left( B\to X_s \gamma \right) = \left( 3.45 \pm 0.15
  \pm 0.40 \right )\times10^{-4}$
where the errors are statistical and systematic, respectively. The
partial branching fraction, mean and variance of the photon energy
spectrum and  the systematic error budget for various lower energy thresholds
are given in Table~\ref{Tab:AVERAGEsummary}~\cite{correlations}.

\begin{table*}
  \caption{\label{Tab:AVERAGEsummary}The measurements of the
    branching fraction, mean and variance of the photon energy
    spectrum for various lower energy thresolds measured in the
    $B$-meson rest frame and the contributions to the systematic uncertainty. 
  }
\begin{ruledtabular}
  \begin{tabular}{|l||rrrr|rrrr|rrrr|}\tiny
    & \multicolumn{4}{c|}{$\mathrm{BF}(B\to X_s\gamma)$ ($10^{-4}$)} &
    \multicolumn{4}{c|}{$\left< E_\gamma \right>$ (GeV)}  &
    \multicolumn{4}{c|}{$\Delta
      E_\gamma^2\equiv\left<E_\gamma^2\right>-\left<E_\gamma\right>^2$ (GeV$^2$)}
    \\ \hline
    $E^{\mathrm{B}}_{\gamma-\mathrm{Low}}$
       [GeV]                   & 1.70&	1.80&	1.90&	2.00&		1.70	&1.80	&1.90	&2.00		&1.70&	1.80	&1.90	&2.00	    \\\hline
Value                          &3.45    &3.36    &3.21    &3.02        &2.282   &2.294   &2.311   &2.334       &0.0428  &0.0370  &0.0302  &0.0230\\
$\pm$statistical               &0.15    &0.13    &0.11    &0.10        &0.015   &0.011   &0.009   &0.007       &0.0047  &0.0029  &0.0019  &0.0014\\
$\pm$systematic                &0.40    &0.25    &0.16    &0.11
    &0.051   &0.028   &0.015   &0.009       &0.0202  &0.0081  &0.0030
    &0.0016\\ \hline
\multicolumn{13}{|c|}{Systematic Uncertainties} \\\hline
1. Continuum           & 0.26 & 0.16 & 0.10 & 0.07 & 0.033 & 0.018 & 0.009 & 0.004 &0.0111 & 0.0048 & 0.0016 & 0.0005   \\
2. Selection           & 0.15 & 0.12 & 0.10 & 0.08 & 0.016 & 0.009 & 0.005 & 0.002 &0.0089 & 0.0029 & 0.0011 & 0.0004   \\
3. $\pi^0/\eta$        & 0.07 & 0.05 & 0.04 & 0.02 & 0.011 & 0.006 & 0.003 & 0.002 &0.0068 & 0.0022 & 0.0007 & 0.0003   \\
4. Other $B$           & 0.25 & 0.14 & 0.06 & 0.02 & 0.033 & 0.017 & 0.007 & 0.002 &0.0121 & 0.0051 & 0.0017 & 0.0004   \\
5. Beam bkgd.          & 0.03 & 0.02 & 0.02 & 0.01 & 0.002 & 0.001 & 0.000 & 0.000 &0.0006 & 0.0003 & 0.0001 & 0.0001   \\
6. Unfolding           & 0.01 & 0.01 & 0.02 & 0.02 & 0.006 & 0.005 & 0.005 & 0.004 &0.0008 & 0.0006 & 0.0005 & 0.0004   \\
7. Model               & 0.01 & 0.01 & 0.00 & 0.01 & 0.002 & 0.001 & 0.000 & 0.001 &0.0010 & 0.0006 & 0.0004 & 0.0004   \\
8. Resolution          & 0.05 & 0.03 & 0.01 & 0.00 & 0.007 & 0.004 & 0.001 & 0.000 &0.0026 & 0.0011 & 0.0004 & 0.0001   \\
9. $\gamma$ Detection  & 0.03 & 0.02 & 0.00 & 0.00 & 0.005 & 0.003 & 0.002 & 0.001 &0.0015 & 0.0007 & 0.0002 & 0.0000   \\
10. $B\to X_d\gamma$    & 0.01 & 0.01 & 0.01 & 0.01 & 0.000 & 0.000 & 0.000 & 0.000 &0.0000 & 0.0000 & 0.0000 & 0.0000   \\
11. Boost               & 0.01 & 0.01 & 0.02 & 0.02 & 0.002 & 0.002 & 0.004 & 0.005 &0.0012 & 0.0005  &0.0008  &0.0009\\
 \end{tabular}
  \end{ruledtabular}    
\end{table*}
\normalsize

In conclusion, for the first time, more than 97\% of the $B\to X_s\gamma$ phase space
is measured~\cite{Buchmuller:2005zv} allowing the theoretical
uncertainties to be significantly reduced. 
The measured branching fractions are in agreement with the latest theoretical 
calculations~\cite{Misiak:2006zs} and are the most precise to date.
Our results
place tighter constraints on models of new
physics~\cite{Buchmueller:2007zk}, where for example, 
in the two-Higgs-doublet model II~\cite{Abbott:1979dt}, the charged Higgs mass is constrained to be above 260 GeV/$c^2$
at the 95\% confidence level~\cite{Misiak:2006zs}. The moment measurements reduce the uncertainty on $|V_{ub}|$. 
Our measurement supersedes our previous result and will be the last of its type from Belle.

We thank the KEKB group for excellent operation of the
accelerator, the KEK cryogenics group for efficient solenoid
operations, and the KEK computer group and
the NII for valuable computing and SINET3 network
support.  We acknowledge support from MEXT, JSPS and Nagoya's TLPRC (Japan);
ARC, DIISR and A.J. Slocum (Australia); NSFC (China); 
DST (India); MOEHRD and KOSEF (Korea); 
MNiSW (Poland); MES and RFAAE (Russia); ARRS (Slovenia); SNSF (Switzerland); 
NSC and MOE (Taiwan); and DOE (USA).

\appendix
\begin{flushleft}
{\bf APPENDIX}

The matrix of the correlation coefficients for the errors of the partial branching
fraction, mean and variance is given in Table~\ref{Tab:Correlations}.
The matrix is symmetric so only the upper half is tabulated. The
moments are measured for four energy ranges, $1.7$,$1.8$,$1.9$,$2.0$
$< E_\gamma < 2.8$ GeV, as measured in the rest frame of the $B$-meson. 
\begin{table*}
  \caption{\label{Tab:Correlations} The correlation coefficients of the
    branching fraction, mean and variance of the photon energy
    spectrum for various lower energy thresolds,
    $E^B_\gamma$, as measured in the rest frame of the $B$-meson.
  }
  \begin{ruledtabular}
    \begin{tabular}{|r||rrrr|rrrr|rrrr|}\tiny
      & \multicolumn{4}{c|}{$\Delta\mathcal{B}$} & \multicolumn{4}{c|}{$\left<E_\gamma \right>$} & \multicolumn{4}{c|}{$\Delta E_\gamma^2$} \\
      &     1.7 &    1.8  &     1.9&    2.0&       1.7 &   1.8 &   1.9 &   2.0 &    1.7   & 1.8   & 1.9  & 2.0\\\hline
      1.7                        &   1.000 &   0.971 &   0.837 &   0.603 &      0.787 &   0.756 &   0.628 &   0.372 &      0.810 &   0.882 &   0.835 &   0.484  \\ 
      1.8                        &         &   1.000 &   0.942 &   0.768 &      0.692 &   0.670 &   0.564 &   0.340 &      0.714 &   0.800 &   0.802 &   0.530  \\ 
      $\Delta\mathcal{B}$ 1.9    &         &         &   1.000 &   0.934 &      0.503 &   0.499 &   0.434 &   0.272 &      0.509 &   0.599 &   0.664 &   0.534  \\ 
      2.0                        &         &         &         &   1.000 &      0.279 &   0.302 &   0.293 &   0.212 &      0.246 &   0.321 &   0.424 &   0.451  \\\hline 
      1.7                        &         &         &         &         &      1.000 &   0.978 &   0.861 &   0.582 &      0.844 &   0.835 &   0.697 &   0.326  \\ 
      1.8                        &         &         &         &         &            &   1.000 &   0.944 &   0.716 &      0.769 &   0.780 &   0.667 &   0.336  \\ 
      $\left<E_\gamma\right>$ 1.9&         &         &         &         &            &         &   1.000 &   0.897 &      0.625 &   0.654 &   0.588 &   0.345  \\ 
      2.0                        &         &         &         &         &            &         &         &   1.000 &      0.388 &   0.440 &   0.458 &   0.384  \\ \hline
1.7                        &         &         &         &         &            &         &         &         &      1.000 &   0.965 &   0.815 &   0.408  \\ 
1.8                        &         &         &         &         &            &         &         &         &            &   1.000 &   0.925 &   0.560  \\ 
$\Delta E_\gamma^2$ 1.9    &         &         &         &         &            &         &         &         &            &         &   1.000 &   0.811  \\ 
2.0                        &         &         &         &         &            &         &         &         &            &         &         &   1.000  \\
 \end{tabular}
  \end{ruledtabular}    
\end{table*}
\normalsize

\end{flushleft}

\end{document}
%